\documentclass[11pt,reqno]{article}
\usepackage{amsfonts}
\usepackage{amsmath}
\usepackage{amsbsy}

\usepackage{amssymb,latexsym}
\numberwithin{equation}{section}

\begin{document}

\title{Gravitational Field Equations in a Braneworld With
Euler-Poincar\'{e} Term}
\begin{titlepage}
\author{ A. N. Aliev\footnote{E.mail: aliev@gursey.gov.tr}
\hspace{0.4cm}, \hspace{0.3cm} H. Cebeci\footnote{E.mail: cebeci@gursey.gov.tr} \\
{\small Feza G\"{u}rsey Institute, 34684 \c{C}engelk\"{o}y,
\.{I}stanbul, Turkey} \\ \\
\\T. Dereli\footnote{E.mail: tdereli@ku.edu.tr}\\{\small
Department of Physics,  Ko\c{c} University, 34450 Sar{\i}yer,
\.{I}stanbul, Turkey} }

\date{ }

\maketitle

\medskip


\begin{abstract}
\noindent  We present the effective gravitational field equations
in a 3-brane world with Euler-Poincar\'{e} term and a cosmological
constant in the bulk spacetime. The similar equations on a 3-brane
with $\mathbb{Z}_{2}$ symmetry embedded in a five dimensional bulk
spacetime were obtained earlier by Maeda and Torii using the
Gauss-Codazzi projective approach in the framework of the Gaussian
normal coordinates. We recover these equations on the brane in
terms of differential forms and using a more general coordinate
setting in the spirit of Arnowitt, Deser and Misner (ADM). The
latter allows for acceleration of the normals to the brane surface
through the lapse function and the shift vector. We show that the
gravitational effects of the bulk space are transmitted to the
brane through the projected ``electric" 1-form field constructed
from the conformal Weyl curvature 2-form of the bulk space. We
also derive the evolution equations into the bulk space for the
electric 1-form field, as well as for the ``magnetic" 2-form field
parts of the bulk Riemann curvature 2-form. As expected, unlike
on-brane equations, the evolution equations involve terms
determined by the nonvanishing acceleration of the normals in the
ADM-type slicing of spacetime.
\end{abstract}
\end{titlepage}

\section{Introduction}
Brane world theories are strictly motivated by the string models.
They were mainly proposed to provide new solutions to hierarchy
problem and compactification of extra dimensions. The main content
of the braneworld idea is that we live in a four dimensional world
embedded in a higher dimensional bulk spacetime. In braneworld
models, it is shown that gauge fields, fermions and scalar fields
of the Standard Model are localised on the brane, while gravity
can freely propagate into the higher dimensional bulk space. Such
a localization can be seen in Horava-Witten model \cite{horava}.
It is based on the idea that strongly coupled 10-dimensional
$E_{8} \times E_{8}$ heterotic string theory can be related to
11-dimensional theory (M-theory) compactified on an
$S^{1}/\mathbb{Z}_{2}$ orbifold with gauge fields propagating on
two 10-dimensional branes located on the boundary hyperplanes with
a $\mathbb{Z}_{2}$ symmetry. It has been shown that the subsequent
compactification of this model on a deformed Calabi-Yau space
leads to a five dimensional spacetime with boundary hyperplanes
becoming two 3-branes describing our 4-dimensional world
\cite{witten,lukas}.

In a model proposed by Arkani-Hamed-Dimopoulos-Dvali (ADD)
\cite{Arkani, Antoniadis}, the idea of a 3-brane universe is
combined with the idea of Kaluza-Klein compactification to solve
the hierarchy problem of high-energy physics. This model considers
our observable four-dimensional spacetime as a 3-brane with all
matter fields localised on it, while gravity can "leak" into all
extra spatial dimensions. The size of the extra dimensions may be
much greater than the conventional Planckian length. Therefore,
unlike the original Kaluza-Klein scenario, the ADD model implies
that the extra dimensions can manifest themselves as physical
ones.

An alternative braneworld model has been proposed by Randall and
Sundrum \cite{randall,sundrum}. Indeed, they proposed two
different models consisting of a single extra spatial dimension.
In their first model, the two 3-branes were located at the
boundaries of an $S^{1}/\mathbb{Z}_{2}$ orbifold in a
5-dimensional Anti-de Sitter ($AdS_{5}$) spacetime. It has been
shown that under a certain fine balance between the self-gravity
of the branes and the bulk cosmological constant the ultraviolet
scale is generated from the large Planckian scale through an
exponential warp function of a small compactification radius. In
the second model, the authors suggested that our observable
universe is of a singe self-gravitating 3-brane with positive
tension and the extra dimension may have even infinite size. In
this case, the fine tuning between the gravitational effects of
the brane and the bulk results in the localization of
5-dimensional graviton zero mode on the 3-brane, while massive
Kaluza-Klein modes die off rapidly at large distances. In other
words, 4-dimensional Newtonian gravity is recovered with high
enough accuracy at low energy scales.

In a covariant approach the localization of gravity on a 3-brane
was studied by Shiromizu, Maeda and Sasaki
\cite{sasaki,shiromizu}. These authors derived covariant
gravitational field equations on a 3-brane embedded in a
five-dimensional bulk spacetime with $\mathbb{Z}_{2}$ symmetry.
Subsequent generalizations of this study can be made by including
dilaton fields as well as the combinations of higher order ($N
\geq 2$) Euler-Poincar\'{e} densities
\cite{lovelock,boulware,zwiebach,zumino}. (See also
\cite{dereli1,dereli2}). In \cite{maeda1,maeda2} the covariant
gravitational field equations on a singular $\mathbb{Z}_{2}$
symmetric 3-brane were generalized by including a combination of
second order ($N=2$) Euler-Poincar\'{e} gravity which is also
known as the Gauss-Bonnet combination. The case of thick branes
when Gauss-Bonnet self-interactions are included was studied in
\cite{gio}. We recall that these type of combinations naturally
arise in the low energy action of superstring theories. The higher
order curvature combinations can help in the resolution of initial
singularities, inflation and fine tuning of cosmological constant
\cite{dereli2,maeda2}.

In papers \cite{shiromizu,maeda2} and other related works, the
authors project the field equations in the bulk spacetime on a
3-brane using the Gauss-Codazzi projective approach with
subsequent specialization to the  Gaussian normal coordinates.
However, the Gaussian normal coordinates imply a very special
slicing of spacetime in the sense that the geodesics orthogonal to
a given hypersurface remain orthogonal to all successive
hypersurfaces in the slicing. This may not affect the field
equations on the brane, since the use of the Gaussian normal
coordinates is at least well justified in a close neighborhood of
the brane. Indeed, in \cite{aliev1} using a more general ADM-type
coordinate setting it has been shown that the effective
gravitational field equations on the 3-brane obtained earlier in
\cite{shiromizu} remain unchanged, however, the evolution
equations off the brane are significantly modified due to the
acceleration of normals to the brane surface in the nongeodesic,
ADM slicing of spacetime.

The use of ADM type coordinates implies the slicing of spacetime
by timelike hypersurfaces pierced by a congruence of spacelike
curves that are not geodesics and do not intersect the
hypersurfaces orthogonally. In other words, in ADM type setting
one can allow for acceleration of the normals to the brane surface
through the lapse function and the shift vector. Therefore, the
consistent analysis of the brane dynamics would benefit from
complete freedom in the slicing of five dimensional spacetime in
the spirit of Arnowitt, Deser and Misner \cite{adm}. For instance,
a black hole on a 3-brane world would have a horizon that extends
into the bulk space (see Ref.\cite{aliev2} for a comprehensive
description of the situation). Clearly, the use of Gaussian normal
coordinates may not be appropriate for this kind of situations and
the ADM type approach may become very important.

In this paper, using the language of differential forms we
generalize the results of \cite{aliev1} by including ($N=2$)
Euler-Poincar\'{e} term in a five-dimensional bulk spacetime. We
show that the gravitational influence of the bulk space is felt on
the brane through the projected ``electric" 1-form field
constructed from the bulk Weyl curvature 2-form. We also derive
the evolution equations into the bulk space for the electric
1-form field, as well as for the ``magnetic" 2-form field parts of
the bulk Riemann curvature 2-form. We show that, unlike the
on-brane equations, the evolution equations are drastically
changed due to the nonvanishing acceleration of the normals in the
ADM-type slicing of spacetime.



\section{Spacetime geometry and gravitational field equations}

We consider a five dimensional bulk spacetime manifold $M$ equipped
with a metric $G$. We suppose that bulk spacetime includes a
3-brane, $4$-dimensional hypersurface endowed with a metric $g$. Our
$5$-dimensional action with a cosmological constant and
Euler-Poincar\'{e} term can be written as
\begin{eqnarray}
S_{5} &=& \int_{M} \left\{ \frac{1}{2} \mathbb{R}^{AB} \wedge \# (
E_{A} \wedge E_{B} ) + \frac{\eta}{4} {\cal L}_{EP} - \Lambda \# 1 \right\}
+ \int_{\partial M} {\cal L} ( Brane ) \nonumber \\
& & + \int_{M} {\cal L} ( Bulk ) .
\label{1}
\end{eqnarray}
Here, the basic gravitational field variables are the coframe 1-forms
$E^{A}$, $\{A = 0,1,2,3,5 \}$, in terms of which the spacetime metric
$G=\eta_{AB} E^{A} \otimes E^{B}$ where $\eta_{AB} = diag ( - ++++)$.
5-dimensional Hodge map $\#$ is defined so that oriented volume form $\#1
= E^{0} \wedge E^{1} \wedge E^{2} \wedge E^{3} \wedge E^{5}$, while
$\Lambda$ is the cosmological constant in the bulk spacetime. The
torsion-free connection 1-forms $ \Omega^{A}\,_{B}$ are obtained from the first
Cartan structure equations
\begin{equation}
d E^{A} + \Omega^{A}\,_{B} \wedge  E^{B} = 0
\label{2}
\end{equation}
where the metric compatibility requires $\Omega_{AB} = - \Omega_{BA} $.
Corresponding curvature 2-forms  are obtained from the second Cartan
structure equations
\begin{equation}
\mathbb{R}^{AB} = d \Omega^{AB} + \Omega^{A}\,_{C} \wedge \Omega^{CB} .
\label{3}
\end{equation}
$N=2$ Euler-Poincar\'{e} form action density \cite{dereli1,dereli2} in
5-dimensional spacetime is
\begin{equation}
{\cal L}_{EP} =  \mathbb{R}^{AB} \wedge \mathbb{R}^{CD} \wedge \# (
E_{A} \wedge E_{B} \wedge E_{C} \wedge E_{D} )
\label{4}
\end{equation}
which can also be written as
\begin{equation}
{\cal L}_{EP} = 2 \mathbb{R}_{AB} \wedge \# \mathbb{R}^{AB} - 4
\mathbb{P}_{A} \wedge \# \mathbb{P}^{A} + {\cal R}_{(5)}^{2} \# 1
\label{5}
\end{equation}
in terms of the 5-dimensional Ricci 1-form $\mathbb{P}^{A} =
\imath_{B}\,\mathbb{R}^{BA}$ and curvature scalar ${\cal R}_{(5)}
= \imath_{A}\,\imath_{B}\,\mathbb{R}^{BA} $. Then the field
equation obtained from the variation of action (\ref{1}) with
respect to co-frame $E^{C}$, is
\begin{eqnarray}
\frac{1}{2} \mathbb{R}^{AB} \wedge \# ( E_{A} \wedge E_{B} \wedge E_{C}
) &=& - \frac{\eta}{4} \mathbb{R}^{AB} \wedge \mathbb{R}^{DG} \wedge \#
( E_{A} \wedge E_{B} \wedge E_{D} \wedge E_{G} \wedge E_{C} ) \nonumber
\\ & & + \Lambda \# E_{C} - \# T_{C} (Bulk) - \# \tau_{C} ( Brane )
\label{6}
\end{eqnarray}
where $\# T_{C}$ is the stress-energy tensor in the bulk space and
$\#\tau_{C}$ is the stress-energy tensor on 3-brane obtained from the
variation of $ \int_{\partial M} {\cal L} (brane)$. We take our
5-dimensional local coordinate chart as $X^{M}:\{ x^{\mu} , y \}$ where $x^{\mu}$
denotes the local coordinates on the brane. We assume that normal to our
brane surface accelerates through the lapse function $\phi(x,y)$ and
the shift vector $N^{a} (x,y)$. Then in terms of the lapse function and
the shift vector the 5-dimensional spacetime metric $G$ can be written
as
\begin{equation}
G = g + N_{\mu} d x^{\mu} \otimes dy + N_{\mu} d y \otimes d x^{\mu} +
( N_{\mu} N^{\mu} + \phi^{2} ) d y \otimes d y
\label{7}
\end{equation}
where $g = g_{\mu \nu} d x^{\mu} \otimes d x^{\nu}$ is the
3-brane spacetime metric. We choose the orthonormal co-frame
1-forms as
\begin{equation}
E^{a} (x,y) = e^{a} (x,y) + N^{a} (x,y) d y \quad a = 0,1,2,3 \qquad
E^{5} = \phi (x,y) d y .
\label{8}
\end{equation}
If we define shift 1-form $N =N_{a} e^{a} $, then $N^{a}$ are the
orthonormal components of the shift vector $\vec N$. By using inner product
identity $\imath_{B}\,E^{A} = \delta_{B}\,^{A}$, we can obtain the
components of corresponding inner product operators $\imath_{A}$. They are
given by
\begin{equation}
\imath_{a} = \iota_{a} , \qquad \imath_{5} = \frac{1}{\phi} \iota_{y} -
\frac{1}{\phi} N^{a} \iota_{a}
\label{9}
\end{equation}
where $\iota_{a}$ are the inner product operators in 4-dimensional
brane spacetime satisfying $\iota_{b}\,e^{a} = \delta_{b}\,^{a} $. We
define $\iota_{y} = \iota_{\frac{\partial}{\partial y}}$ shortly.
From (\ref{2}), we can determine the torsion-free connection 1-forms
$\Omega^{A}\,_{B}$ as
\begin{equation}
2 \Omega^{A B} = \imath^{B} d E^{A} - \imath^{A} d E^{B} + ( \imath^{A}
\imath^{B} d E_{C} ) E^{C} .
\label{10}
\end{equation}
On the other hand, we can write $e^{a} = e^{a}\,_{\mu} (x,y)\,d
x^{\mu}$ in terms of local coordinate basis one-forms $d x^{\mu}$. Then since
$e^{a} = e^{a} (x,y)$, we have the following equation satisfied by
torsion-free connection 1-forms $\omega^{a}\,_{b}$ and orthonormal
co-frames $e^{a}$ of 3-brane spacetime:
\begin{equation}
d e^{a} = H^{a}\,_{b} dy \wedge e^{b} - \omega^{a}\,_{b} \wedge e^{b} ,
\label{11}
\end{equation}
where
\begin{equation}
H^{a}\,_{b} = \partial_{y} e^{a}\,_{\mu} \tilde{e}^{\mu}\,_{b} .
\label{12}
\end{equation}
We note that $e^{a}\,_{\mu}$ and $\tilde{e}^{\mu}\,_{b}$ satisfy the
orthogonality relation $ e^{a}\,_{\mu}\, \tilde{e}^{\mu}\,_{b} =
\delta^{a}\,_{b} $.
Then using (\ref{10}), we calculate the components of connection
1-forms $\Omega^{A}\,_{B}$ as
\begin{equation}
\Omega^{ab} = \omega^{ab} + \frac{1}{\phi} \lambda^{ab} E^{5}
\label{13}
\end{equation}
where
\begin{equation}
\lambda_{ab} = \frac{1}{2} ( H_{ba} - H_{ab} + \iota_{b} D N_{a} -
\iota_{a} D N_{b} )
\label{14}
\end{equation}
and
\begin{equation}
\Omega^{5 a} = - K^{a}\,_{b} e^{b} - \frac{1}{\phi} \left( K^{a}\,_{b}
N^{b} - \partial^{a} \phi \right) E^{5}
\label{15}
\end{equation}
where we have introduced the quantity
\begin{equation}
K_{ab} = \frac{1}{2 \phi} \left\{ ( H_{ab} + H_{ba} ) - ( \iota_{a} D
N_{b} + \iota_{b} D N_{a} ) \right\}
\label{16}
\end{equation}
as an extrinsic curvature term and $ K = \eta_{ab} K^{ab} $. We also
note that we can write connection 1-forms $\omega^{ab}$ of brane
spacetime as $ \omega^{ab} = \omega_{\mu}\,^{ab} d x^{\mu} $. Then since
$\omega^{ab} = \omega^{ab} (x,y)$, we have the following equation satisfied
by the curvature 2-forms $R^{ab}$ and connection 1-forms $\omega^{ab}$
on the brane
\begin{equation}
d \omega^{ab} = H^{ab}\,_{c} d y \wedge e^{c} - \omega^{a}\,_{c} \wedge
\omega^{cb} + R^{ab}
\label{17}
\end{equation}
where
\begin{equation}
H^{ab}\,_{c} = \partial_{y} \omega_{\mu}\,^{ab} \tilde{e}_{c}\,^{\mu} .
\label{18}
\end{equation}
Next, using (\ref{3}), we calculate the orthonormal components of
curvature 2-forms
\begin{equation}
\mathbb{R}^{ab} = \pi^{ab} + \tau^{ab} \wedge E^{5} .
\label{19}
\end{equation}
Here
\begin{equation}
\pi^{ab} = R^{ab} - K^{a}\,_{c} K^{b}\,_{d} e^{c} \wedge e^{d} ,
\label{20}
\end{equation}
and
\begin{eqnarray}
\tau^{ab} &=& \frac{1}{\phi} \left\{ D \lambda^{ab} + K^{a}\,_{c}
K^{b}\,_{d} N^{c} e^{d} - K^{a}\,_{c} K^{b}\,_{d} N^{d} e^{c} \right.
\nonumber \\
& & \left. + K^{a}\,_{c} \partial^{b} \phi e^{c} - \partial^{a} \phi
K^{b}\,_{d} e^{d} - H^{ab}\,_{c} e^{c} \right\}
\label{21}
\end{eqnarray}
where $R^{ab}$ are the curvature 2-forms on the brane with connection
$\omega^{a}\,_{b}$. We also calculate the remaining curvature components
as
\begin{equation}
\mathbb{R}^{5 a} = \rho^{a} + \sigma^{a} \wedge E^{5}
\label{22}
\end{equation}
where
\begin{equation}
\rho^{a} = - D ( K^{a}\,_{b} ) \wedge e^{b}
\label{23}
\end{equation}
and
\begin{eqnarray}
\sigma^{a} &=& \frac{1}{\phi} \left\{ \partial_{y} K^{a}\,_{b} e^{b} -
D ( K^{a}\,_{b} N^{b} )  + D ( \partial^{a} \phi ) \right. \nonumber \\
& & \left. - \lambda^{ca} K_{cb} e^{b} + K^{a}\,_{c} H^{c}\,_{b} e^{b}
\right\} .
\label{24}
\end{eqnarray}
By using equations (\ref{19}) and (\ref{22}), we can calculate the
components of Ricci 1-form $\mathbb{P}^{A}$ as
\begin{equation}
\mathbb{P}^{a} = \iota_{b} \pi^{ba} - \frac{1}{\phi} N_{b} \iota^{b}
\rho^{a} - \sigma^{a} + \left( \iota_{b} \tau^{ba} - \frac{1}{\phi} N_{b}
\iota^{b} \sigma^{a} \right) E^{5}
\label{25}
\end{equation}
and
\begin{equation}
\mathbb{P}^{5} = - \iota_{b} \rho^{b} - \iota_{b} \sigma^{b} E^{5} .
\label{26}
\end{equation}
Using $ {\cal R}_{(5)} = \imath_{A} \imath_{B} \mathbb{R}^{BA} $,
we can calculate the curvature scalar $ {\cal R}_{(5)} $ of bulk
spacetime in terms of curvature scalar $ {\cal R}_{(4)} $ of brane
spacetime defined by $ {\cal R}_{(4)} = \iota_{a} \iota_{b} R^{ba}
$:
\begin{eqnarray}
{\cal R}_{(5)} &=& {\cal R}_{(4)} - K^{2} + K_{ab} K^{ab} +
\frac{2}{\phi} \left\{ N_{b} \iota^{b} D K - \iota_{a} D( \partial^{a} \phi )
\right. \nonumber \\
& & \left. + \iota_{a} D N^{b} K^{a}\,_{b} - \partial_{y} K -
H^{b}\,_{a} K^{a}\,_{b} \right\} .
\label{27}
\end{eqnarray}
Next, using the relation
\begin{equation}
\iota_{a} D N_{b} - H_{ba} = \lambda_{ba} - \phi K_{ba}
\label{28}
\end{equation}
and the symmetry of $K_{ba}$ and the anti-symmetry of
$\lambda_{ba}$, we can simplify equation (\ref{27}) as
\begin{equation}
{\cal R}_{(5)} = {\cal R}_{(4)} - K^{2} - K_{ab} K^{ab} +
\frac{2}{\phi} \left\{ {\cal L}_{\vec N} K - \partial_{y} K - \iota_{a} D (
\partial^{a} \phi ) \right\}
\label{29}
\end{equation}
where
\begin{equation}
{\cal L}_{\vec N} K = N^{b} \iota_{b} D K
\label{30}
\end{equation}
is the Lie derivative of $K$ along the vector $\vec N$. In order
to obtain the gravitational field equations on the brane, we note
the following identities between 5-dimensional Hodge map $\#$ and
4-dimensional Hodge map $\ast$:
\begin{equation}
\begin{array}{c}
\# 1 = \ast 1 \wedge E^{5} , \\
\# E^{a} = \ast e^{a} \wedge E^{5} , \\
\# E^{5} = \ast 1 - \frac{N^{a}}{\phi} \ast e_{a} \wedge E^{5} , \\
\# ( E_{a} \wedge E_{b} \wedge E_{c} ) = \ast ( e_{a} \wedge e_{b}
\wedge e_{c} ) \wedge E^{5} , \\
\# ( E_{b} \wedge E_{c} \wedge E_{5} ) = \ast ( e_{b} \wedge e_{c} ) -
\frac{N^{a}}{\phi} \ast ( e_{b} \wedge e_{c} \wedge e_{a} ) \wedge
E^{5} .
\end{array}
\label{31}
\end{equation}
We substitute (\ref{19}) and (\ref{22}) into field equation
(\ref{6}) and obtain reduced gravitational field equations on the
3-brane. For $C=c$, we obtain the field equations,
\begin{eqnarray}
\frac{1}{2} \pi^{ab} \wedge \ast ( e_{a} \wedge e_{b} \wedge e_{c} ) -
\frac{1}{\phi} N^{d} \rho^{b} \wedge \ast ( e_{b} \wedge e_{c} \wedge
e_{d} ) \nonumber \\
+ \sigma^{b} \wedge \ast ( e_{b} \wedge e_{c} ) = \eta \left\{ \pi^{ab}
\wedge \sigma^{d} + \tau^{ab} \wedge \rho^{d} \right\} \epsilon_{abcd}
\nonumber \\
+ \Lambda \ast e_{c} - T_{cd} \ast e^{d} + \frac{1}{\phi} T_{c5} N_{a}
\ast e^{a} - \tau_{cd} \ast e^{d}
\label{32}
\end{eqnarray}
and
\begin{eqnarray}
D K^{b}\,_{d} \wedge e^{d} \wedge \ast ( e_{b} \wedge e_{c} ) &=& \eta
\left\{ R^{ab} - K^{a}\,_{s} K^{b}\,_{u} e^{s} \wedge e^{u} \right\}
\wedge D K^{d}\,_{l} \wedge e^{l} \epsilon_{abcd} \nonumber \\
& &  + T_{c5} \ast 1 .
\label{33}
\end{eqnarray}
We see that (\ref{33}) is the momentum constraint equation with
Euler-Poincar\'{e} term, $T_{cd}$ is the energy-momentum tensor of bulk
spacetime projected on the brane surface, while $T_{c5} = J_{c}$ describes
the momentum flux from the brane or onto the brane. We note that,
\begin{equation}
\# \tau_{c} = \tau_{cd} \ast e^{d} \wedge E^{5}
\label{34}
\end{equation}
where the stress-energy tensor on the brane is of the form
\begin{equation}
\tau_{cd} = \tilde{\tau}_{cd} \frac{\delta(y)}{\phi}
\label{35}
\end{equation}
locating the brane hypersurface at $y=0$. We relate $\delta$-function
behaviour in brane energy-momentum tensor on the right hand side of
equation (\ref{32}) to the jump in the extrinsic curvature terms of brane
\cite{davis,gravanis}. At this stage, we impose $\mathbb{Z}_{2}$ mirror
symmetry and integrate equation (\ref{32}) across the brane surface
along the orbits of evolution vector $Z^{A}$. We assume that quantities
$K^{+}_{ab}$ and $K^{-}_{ab}$ evaluated on both sides of the brane,
respectively, remain bounded. Then we obtain the following junction
conditions
\begin{equation}
e^{+}_{a} = e^{-}_{a}
\label{36}
\end{equation}
and
\begin{eqnarray}
[K_{ac}]^{+}_{-} - [K]^{+}_{-} \eta_{ac} - \eta \left\{ 2 R_{abcd}
[K^{db}]^{+}_{-} + 2 P_{ab} [K^{b}\,_{c}]^{+}_{-} - 2 P_{ac} [K]^{+}_{-}
\right. \nonumber \\
 \left. + 2 P_{cb} [K_{a}\,^{b}]^{+}_{-} - {\cal R}_{(4)}
[K_{ca}]^{+}_{-}
 + ( {\cal R}_{(4)} [K]^{+}_{-} - 2 P_{bd} [K^{db}]^{+}_{-} ) \eta_{ca}
\right\} \nonumber \\
+ \eta \left\{ 2 [K K^{b}\,_{c} K_{ba} ]^{+}_{-} - 2 [K_{as} K^{sb}
K_{bc}]^{+}_{-}
 + [K_{bs}K^{sb} K_{ca}]^{+}_{-} - [K^{2} K_{ca}]^{+}_{-}  \right.
\nonumber \\
\left. + \left( \frac{2}{3} [K^{s}\,_{b} K^{b}\,_{n} K^{n}\,_{s}
]^{+}_{-}
- [K^{bd} K_{db} K]^{+}_{-} + \frac{1}{3} [K^{3}]^{+}_{-} \right)
\eta_{ca} \right\} = - \tilde{\tau}_{ca}
\label{37}
\end{eqnarray}
where we have defined the components of the Riemann tensor
$R_{abcd}$ and the  Ricci tensor $P_{ab}$ from the relations
$R_{ab} = \frac{1}{2} R_{abcd} e^{c} \wedge e^{d} $ and $P_{a} =
P_{ab} e^{b}$. Due to $\mathbb{Z}_{2}$-symmetry, $K_{ab}^{+} = -
K_{ab}^{-}$. Using this fact and dropping the $\pm$ indices, we
obtain the equation that relates stress-energy tensor of the brane
to the extrinsic curvature terms $K_{ab}$ and the intrinsic
curvature terms on either side of the brane surface:
\begin{eqnarray}
& &K_{ac} - K \eta_{ac} - \eta \left\{ 2 R_{abcd} K^{db} + 2 P_{ab}
K^{b}\,_{c} - 2 P_{ac} K + 2 P_{cb} K_{a}\,^{b} - {\cal R}_{(4)} K_{ca}
\right. \nonumber \\
& &\left. + ( {\cal R}_{(4)} K - 2 P_{sb} K^{bs} ) \eta_{ca} \right\} +
\eta \left\{ 2 K K^{b}\,_{c} K_{ba} - 2 K_{as} K^{sb} K_{bc} - K^{2}
K_{ca} \right. \nonumber \\
& & \left. + K_{bd} K^{db} K_{ca} + ( \frac{2}{3} K^{s}\,_{b}
K^{b}\,_{d} K^{d}\,_{s} - K_{bn} K^{nb} K + \frac{1}{3} K^{3} ) \eta_{ca}
\right\} = - \frac{1}{2} \tilde{\tau}_{ca}  .
\label{38}
\end{eqnarray}
Next, we obtain the reduced field equations for $C=5$:
\begin{eqnarray}
\frac{1}{2} \tau^{ab} \wedge \ast (e_{a} \wedge e_{b} ) - \frac{1}{2
\phi} N^{d} \pi^{ab} \wedge \ast  ( e_{a} \wedge e_{b} \wedge e_{d} ) = -
\frac{\eta}{2} \pi^{ab} \wedge \tau^{cd} \epsilon_{abcd} \nonumber \\ -
\frac{1}{\phi} \Lambda N_{a} \ast e^{a} - T_{5d} \ast e^{d} +
\frac{1}{\phi} T_{5}\,^{5} N_{a} \ast e^{a}  ,
\label{39}
\end{eqnarray}
\begin{eqnarray}
\frac{1}{2} \left\{ R^{ab} - K^{a}\,_{c} K^{b}\,_{d} e^{c} \wedge e^{d}
\right\} \wedge \ast ( e_{a} \wedge e_{b} ) = - \frac{\eta}{4} \left\{
R^{ab} \wedge R^{cd} \epsilon_{abcd} \right. \nonumber \\ \left. - 2
R^{ab} \wedge K^{c}\,_{s} K^{d}\,_{u} e^{s} \wedge e^{u} \epsilon_{abcd}
\right. \nonumber \\ \left. + K^{a}\,_{l} K^{b}\,_{n} K^{c}\,_{s}
K^{d}\,_{u} e^{l} \wedge e^{n} \wedge e^{s} \wedge e^{u} \epsilon_{abcd}
\right\} \nonumber \\
+ \Lambda \ast 1 - T_{5}\,^{5} \ast 1 .
\label{40}
\end{eqnarray}
The expression (\ref{40}) is the Hamiltonian constraint equation with
Euler-Poincar\'{e} term. The component $T_{5}\,^{5} \equiv P$ can be
interpreted as some kind of pressure term.
Next, we decompose the bulk curvature 2-form $\mathbb{R}^{5a}$ into its
``electric" and ``magnetic" parts. Writing $\mathbb{R}^{5a}$ in the
form
\begin{equation}
\mathbb{R}^{5a} = \frac{1}{2} B^{a}\,_{bc} E^{b} \wedge E^{c} +
\tilde{\sigma}^{a}\,_{b} E^{5} \wedge E^{b} ,
\label{41}
\end{equation}
we can obtain the magnetic tensor component
\begin{equation}
B^{a}\,_{bc} = \rho^{a}\,_{bc} = \iota_{c} D K^{a}\,_{b} - \iota_{b} D
K^{a}\,_{c} ,
\label{42}
\end{equation}
and the electric tensor component
\begin{eqnarray}
\tilde{\sigma}^{a}\,_{b} &=& \frac{1}{\phi} \left\{ {\cal L}_{\vec N}
K^{a}\,_{b} - \partial_{y} K^{a}\,_{b} + K_{sb} H^{as} - H_{sb} K^{sa} -
\iota_{b} D ( \partial^{a} \phi ) \right. \nonumber \\
& & \left. - \phi K_{sb} K^{sa} \right\}
\label{43}
\end{eqnarray}
where
\begin{equation}
{\cal L}_{\vec N} K^{a}\,_{b} = N^{s} \iota_{s} D K^{a}\,_{b} +
K^{a}\,_{s} \iota_{b} D N^{s} - K^{s}\,_{b} \iota_{s} D N^{a}
\label{44}
\end{equation}
is the Lie derivative of $K^{a}\,_{b}$ along the vector $\vec N$.
From (\ref{42}), we let the magnetic 2-forms to be given by
\begin{equation}
B^{a} = \frac{1}{2} B^{a}\,_{bc} e^{b} \wedge e^{c} = \rho^{a} .
\label{45}
\end{equation}
We note that the trace of the magnetic tensor is not zero, i.e
$\iota_{a} \rho^{a} \not = 0 $. On the other hand from (\ref{43}),
we let the electric 1-form to be given by
\begin{equation}
\tilde{\sigma}^{a} = \tilde{\sigma}^{a}\,_{b} e^{b} .
\label{46}
\end{equation}
We note that the following relation between $\tilde{\sigma}^{a}$ and
$\sigma^{a}$ holds:
\begin{eqnarray}
\tilde{\sigma}^{a} &=& \frac{1}{\phi} \left\{ N^{s} \iota_{s} D
K^{a}\,_{b} e^{b} - N^{b} D K^{a}\,_{b} \right\} - \sigma^{a} \nonumber \\
&=& - \frac{N^{b}}{\phi} \iota_{b} \rho^{a} - \sigma^{a} .
\label{47}
\end{eqnarray}
From equation (\ref{43}) or (\ref{47}), we can calculate the trace
$\tilde{\sigma}$ of the electric tensor as
\begin{equation}
\tilde{\sigma} = \iota_{a} \tilde{\sigma}^{a} = \left\{ {\cal L}_{\vec
N} K - \partial_{y} K - \phi K_{ab} K^{ab} - \iota_{a} D ( \partial^{a}
\phi ) \right\} \frac{1}{\phi} .
\label{48}
\end{equation}
By comparing equations (\ref{29}) and (\ref{48}), we obtain the
following equation between the trace $\tilde{\sigma}$ and the
curvature scalars ${\cal R}_{(4)}$ and ${\cal R}_{(5)}$:
\begin{equation}
\tilde{\sigma} = \frac{1}{2} \left\{ {\cal R}_{(5)} - ( {\cal R}_{(4)}
- K^{2} + K_{ab} K^{ab} ) \right\} . \label{49}
\end{equation}
On the other hand, we take the trace of field equation (\ref{6}) by
considering its exterior product with $E^{C}$ and obtain the curvature
scalar ${\cal R}_{(5)}$ as
\begin{eqnarray}
\frac{3}{2} {\cal R}_{(5)} \ast 1 &=& - \frac{\eta}{4} \pi^{ab} \wedge
\pi^{cd} \epsilon_{abcd} - \eta \left\{ \pi^{ab} \wedge \sigma^{d}
\wedge e^{c} + \tau^{ab} \wedge \rho^{d} \wedge e^{c} \right. \nonumber \\
& & \left. - \frac{N^{c}}{\phi} \pi^{ab} \wedge \rho^{d} \right\}
\epsilon_{abcd} - T \ast 1 + 5 \Lambda \ast 1
\label{50}
\end{eqnarray}
where
\begin{equation}
T = T_{cd} \eta^{cd} + P .
\label{51}
\end{equation}
We can express equation (\ref{50}) in terms of the electric 1-form
$ \tilde{\sigma}^{a} $ and the magnetic 2-form $\rho^{a}$. Before
that, we simplify $\tau^{ab}$ as
\begin{equation}
\tau^{ab} = - \rho_{s}\,^{ab} e^{s} - \frac{N^{s}}{\phi} \iota_{s}
\pi^{ab}
\label{52}
\end{equation}
where
\begin{equation}
\rho_{s}\,^{ab} = \iota^{b} D K_{s}\,^{a} - \iota^{a} D K_{s}\,^{b} .
\label{53}
\end{equation}
Then using (\ref{52}), equation (\ref{50}) becomes
\begin{eqnarray}
\frac{3}{2} {\cal R}_{(5)} \ast 1 &=& - \frac{\eta}{4} \pi^{ab} \wedge
\pi^{cd} \epsilon_{abcd} + \eta \pi^{ab} \wedge \tilde{\sigma}^{d}
\wedge e^{c} \epsilon_{abcd} \nonumber \\
& & + \eta \rho_{s}\,^{ab} e^{s} \wedge \rho^{d} \wedge e^{c}
\epsilon_{abcd} - T \ast 1 + 5 \Lambda \ast 1 .
\label{54}
\end{eqnarray}
Furthermore, using momentum constraint equation (\ref{33}) and equation
(\ref{52}), we can also express our reduced gravitational field
equation (\ref{32}) in terms of $\tilde{\sigma}^{a}$ and $\rho^{a}$
approaching the brane from either positive $(+)$ or $(-)$ side:
\begin{eqnarray}
\frac{1}{2} \pi^{ab} \wedge \ast ( e_{a} \wedge e_{b} \wedge e_{c} ) -
\tilde{\sigma}^{b} \wedge \ast ( e_{b} \wedge e_{c} ) &=& - \eta
\pi^{ab} \wedge \tilde{\sigma}^{d} \epsilon_{abcd} - \eta \rho_{s}\,^{ab}
e^{s} \wedge \rho^{d} \epsilon_{abcd} \nonumber \\
& & - T_{cd} \ast e^{d} + \Lambda \ast e_{c} .
\label{55}
\end{eqnarray}
Now, we can define the trace-free electric 1-form
$\bar{\sigma}^{a}$ (i.e $\iota_{a} \bar{\sigma}^{a} = 0 $ ) in
terms of $\tilde{\sigma}^{a}$ and $\tilde{\sigma}$ as
\begin{equation}
\bar{\sigma}^{a} = \tilde{\sigma}^{a} - \frac{1}{4} \tilde{\sigma}
e^{a} .
\label{56}
\end{equation}
Then in terms of $\bar{\sigma}^{a}$, the effective gravitational field
equation (\ref{55}) can be written as
\begin{eqnarray}
\frac{1}{2} \pi^{ab} \wedge \ast ( e_{a} \wedge e_{b} \wedge e_{c} ) -
\bar{\sigma}^{b} \wedge \ast ( e_{b} \wedge e_{c} ) + \frac{3}{4}
\tilde{\sigma} \ast e_{c} = - \eta \pi^{ab} \wedge \bar{\sigma}^{d}
\epsilon_{abcd} \nonumber \\
- \frac{\eta}{4} \tilde{\sigma} \pi^{ab} \wedge \ast ( e_{a} \wedge
e_{b} \wedge e_{c} ) - \eta \rho_{s}\,^{ab} e^{s} \wedge \rho^{d}
\epsilon_{abcd} - T_{cd} \ast e^{d} + \Lambda \ast e_{c}
\label{57}
\end{eqnarray}
where
\begin{eqnarray}
\tilde{\sigma} = \frac{1}{3+ \frac{\eta}{2} \tilde{\pi} } \left\{
\frac{\eta}{4} \ast ( \pi^{ab} \wedge \pi^{cd} ) \epsilon_{abcd} - \eta \ast
( \pi^{ab} \wedge \bar{\sigma}^{d} \wedge e^{c} ) \epsilon_{abcd}
\right. \nonumber \\
\left. - \eta \rho_{s}\,^{ab} \ast ( e^{s} \wedge \rho^{d} \wedge e^{c}
) \epsilon_{abcd} - T - \frac{3}{2} \tilde{\pi} + 5 \Lambda \right\}
\label{58}
\end{eqnarray}
in terms of $\bar{\sigma}^{a}$ and we define $ \tilde{\pi} =
\iota_{a}\,\iota_{b}\,\pi^{ba} $. Explicitly, $ \tilde{\pi} $ can be written as
\begin{equation}
\tilde{\pi} = {\cal R}_{(4)} - K^{2} + K_{ab} K^{ab} .
\label{59}
\end{equation}
Next, we consider the Weyl (conformal) curvature 2-form
\begin{equation}
\mathbb{C}^{AB} = \mathbb{R}^{AB} - \frac{1}{3} ( E^{A} \wedge
\mathbb{P}^{B} - E^{B} \wedge \mathbb{P}^{A} ) + \frac{1}{12} E^{A} \wedge
E^{B} {\cal R}_{(5)} .
\label{60}
\end{equation}
Writing the bulk conformal 2-form $\mathbb{C}^{5a}$ as
\begin{equation}
\mathbb{C}^{5a} = \frac{1}{2} M^{a}\,_{bc} E^{b} \wedge E^{c} + {\cal
E}^{a}\,_{b} E^{5} \wedge E^{b} ,
\label{61}
\end{equation}
we can obtain the conformal magnetic tensor component
\begin{equation}
M^{a}\,_{bc} = \rho^{a}\,_{bc} - ( \rho^{s}\,_{sc} \eta^{a}\,_{b} -
\rho^{s}\,_{sb} \eta^{a}\,_{c} ) \frac{1}{3} .
\label{62}
\end{equation}
from which the conformal magnetic 2-form $M^{a}$ follows as
\begin{equation}
M^{a} = \frac{1}{2} M^{a}\,_{bc} e^{b} \wedge e^{c} = \rho^{a} +
\frac{1}{3} \iota_{b} \rho^{b} \wedge e^{a} .
\label{63}
\end{equation}
We note that $M^{a}\,_{bc}$ is traceless, i.e $ \iota_{a} M^{a} =
0 $. On the other hand, we obtain the conformal electric tensor
component in the form
\begin{equation}
{\cal E}^{a}\,_{b} = \frac{2}{3} \tilde{\sigma}^{a}\,_{b} - \frac{1}{3}
\pi^{sa}\,_{sb} - \frac{1}{3} \tilde{\sigma} \eta^{a}\,_{b} +
\frac{1}{12} \eta^{a}\,_{b} {\cal R}_{(5)} .
\label{64}
\end{equation}
If one defines the conformal electric 1-form ${\cal E}^{a} = {\cal
E}^{a}\,_{b} e^{b} $, then
\begin{equation}
{\cal E}^{a} = \frac{2}{3} \bar{\sigma}^{a} + \frac{1}{12} \tilde{\pi}
e^{a} - \frac{1}{3} \iota_{b} \pi^{ba}
\label{65}
\end{equation}
being expressed in terms of the traceless electric 1-forms
$\bar{\sigma}^{a} $. Therefore, in terms of the conformal electric
1-form ${\cal E}^{a}$, the effective field equation (\ref{57})
becomes
\begin{eqnarray}
\frac{1}{2} \pi^{ab} \wedge \ast ( e_{a} \wedge e_{b} \wedge e_{c} ) -
\frac{3}{2} {\cal E}^{b} \wedge \ast ( e_{b} \wedge e_{c} ) - \left(
\frac{9 \tilde{\pi}}{8 \left( 3 + \frac{\eta}{2} \tilde{\pi} \right)} -
\frac{3}{4} \tilde{e} \right) \ast e_{c} \nonumber \\ - \frac{1}{2}
\iota_{s} \pi^{sb} \wedge \ast ( e_{b} \wedge e_{c} ) - \frac{3 \eta}{8
\left( 3 + \frac{\eta}{2} \tilde{\pi} \right) } \kappa \ast e_{c} = -
\frac{3}{2} \eta \pi^{ab} \wedge {\cal E}^{d} \epsilon_{abcd} \nonumber \\
- \left( \frac{\eta}{4} \tilde{e} - \frac{3 \eta \tilde{\pi}}{8 \left(
3 + \frac{\eta}{2} \tilde{\pi} \right)} \right) \pi^{ab} \wedge \ast (
e_{a} \wedge e_{b} \wedge e_{c} )
 - \frac{1}{2} \eta \pi^{ab} \wedge \iota_{n} \pi^{nd} \epsilon_{abcd}
\nonumber \\ + \frac{\eta^{2}}{8 \left( 3 + \frac{\eta}{2} \tilde{\pi}
\right) } \kappa \pi^{ab} \wedge \ast ( e_{a} \wedge e_{b} \wedge e_{c}
) - \eta \rho_{s}\,^{ab} e^{s} \wedge \rho^{d} \epsilon_{abcd} +
\Lambda \ast e_{c} \nonumber \\
- T_{cd} \ast e^{d}
\label{66}
\end{eqnarray}
where
\begin{equation}
\kappa = \ast ( \pi^{sn} \wedge \iota_{k} \pi^{km} \wedge e^{l} )
\epsilon_{snlm}
\label{67}
\end{equation}
and
\begin{eqnarray}
\tilde{e} &=& \frac{1}{ 3 + \frac{\eta}{2} \tilde{\pi}} \left\{
\frac{\eta}{4} \ast ( \pi^{ab} \wedge \pi^{cd} ) \epsilon_{abcd} - \frac{3
\eta }{2} \ast ( \pi^{ab} \wedge {\cal E}^{d} \wedge e^{c} )
\epsilon_{abcd} \right. \nonumber \\
& & \left. - \eta \rho_{s}\,^{ab} \ast ( e^{s} \wedge \rho^{d} \wedge
e^{c} ) \epsilon_{abcd} - T - \frac{3}{2} \tilde{\pi} + 5 \Lambda
\right\} .
\label{68}
\end{eqnarray}
Finally, we note that the effective gravitational field equations
do not explicitly involve acceleration terms determined through
the derivative of the lapse function $\phi$ and the shift vector
$N^{a}$. The effective gravitational field equations on the brane
can be described in terms of either electric 1-form coming from
bulk Riemann curvature 2-form $\mathbb{R}^{5a}$ or conformal
electric 1-form coming from bulk conformal curvature 2-form
$\mathbb{C}^{5a}$.
\section{Evolution equations}
We note that the effective gravitational field equations on the
brane are not closed. To obtain a closed system, we should derive
the off-brane evolution equations. These equations are found by
considering the 5-dimensional Bianchi identity
\begin{equation}
\mathbb{D} \mathbb{R}^{AB} = 0 .
\label{69}
\end{equation}
From
\begin{equation}
\mathbb{D} \mathbb{R}^{ab} = 0 ,
\label{70}
\end{equation}
we obtain
\begin{equation}
D \pi^{ab} + K^{a}\,_{c} e^{c} \wedge \rho^{b} - K^{b}\,_{c} e^{c}
\wedge \rho^{a} = 0
\label{71}
\end{equation}
and
\begin{eqnarray}
\frac{1}{2} \partial_{y} \pi^{ab}\,_{cd} e^{c} \wedge e^{d} +
\frac{1}{2} \pi^{ab}\,_{cd} H^{c}\,_{l} e^{l} \wedge e^{d} + \frac{1}{2}
\pi^{ab}\,_{cd} H^{d}\,_{n} e^{c} \wedge e^{n} \nonumber \\
+ d \phi \wedge \tau^{ab} + \phi D \tau^{ab} + \lambda^{a}\,_{c}
\pi^{cb} + \phi K^{a}\,_{c} e^{c} \wedge \sigma^{b} \nonumber \\
+ K^{a}\,_{c} N^{c} \rho^{b} - \partial^{a} \phi \rho^{b} +
\lambda^{b}\,_{d} \pi^{ad} - \phi K^{b}\,_{c} e^{c} \wedge \sigma^{a} \nonumber \\
- K^{b}\,_{c} N^{c} \rho^{a} + \partial^{b} \phi \rho^{a} = 0 .
\label{72}
\end{eqnarray}
From equation (\ref{71}), we obtain the 4-dimensional Bianchi identity
\begin{equation}
D R^{ab} = 0 .
\label{74}
\end{equation}
Then using equations (\ref{47}) and (\ref{52}) we can simplify
(\ref{72}) as
\begin{eqnarray}
\frac{1}{2} \partial_{y} \pi^{ab}\,_{cd} e^{c} \wedge e^{d} - {\cal
L}_{\vec N} \pi^{ab} + \frac{1}{2} \pi^{ab}\,_{cd} H^{c}\,_{l} e^{l}
\wedge e^{d} + \frac{1}{2} \pi^{ab}\,_{cd} H^{d}\,_{n} e^{c} \wedge e^{n}
\nonumber \\
- \rho_{s}\,^{ab} d \phi \wedge e^{s} - \phi D ( \rho_{s}\,^{ab} )
\wedge e^{s} + \lambda^{a}\,_{c} \pi^{cb} - \phi K^{a}\,_{c} e^{c} \wedge
\tilde{\sigma}^{b} + \phi K^{b}\,_{c} e^{c} \wedge \tilde{\sigma}^{a}
\nonumber \\
- \partial^{a} \phi \rho^{b} + \lambda^{b}\,_{d} \pi^{ad} +
\partial^{b} \phi \rho^{a} = 0
\label{75}
\end{eqnarray}
where
\begin{equation}
{\cal L}_{\vec N} \pi^{ab} = D ( N^{s} \iota_{s} \pi^{ab} ) + N^{s}
\iota_{s} D \pi^{ab}
\label{76}
\end{equation}
is Lie derivative of $\pi^{ab}$ along $\vec N$. We note that the
components of the acceleration of normal to the brane surface are
$ a^{b} = - \frac{1}{\phi}
\partial^{b} \phi $ and $ a^{5} = - \frac{1}{\phi} \partial_{a} \phi
N^{a} $ where $ \partial_{b} \phi = \iota_{b} d \phi $. On the other
hand, from
\begin{equation}
\mathbb{D} \mathbb{R}^{5a} = 0 ,
\label{77}
\end{equation}
we obtain
\begin{equation}
D \rho^{a} - K_{cs} e^{s} \wedge \pi^{ca} = 0
\label{78}
\end{equation}
and
\begin{eqnarray}
\frac{1}{2} \partial_{y} \rho^{a}\,_{bc} e^{b} \wedge e^{c} +
\frac{1}{2} \rho^{a}\,_{bc} H^{b}\,_{l} e^{l} \wedge e^{c} + \frac{1}{2}
\rho^{a}\,_{bc} H^{c}\,_{s} e^{b} \wedge e^{s} + d \phi  \wedge \sigma^{a}
\nonumber \\
+ \phi\,D \sigma^{a} - \phi K_{cs} e^{s} \wedge \tau^{ca} - K_{cs}
N^{s} \pi^{ca}  + \partial_{c} \phi \pi^{ca} + \lambda^{a}\,_{b} \rho^{b} =
0 .
\label{79}
\end{eqnarray}
Using equations (\ref{47}), (\ref{52}) and (\ref{78}), we can write
equation (\ref{79}) as
\begin{eqnarray}
\frac{1}{2} \partial_{y} \rho^{a}\,_{bc} e^{b} \wedge e^{c} - {\cal
L}_{\vec N} \rho^{a} + \frac{1}{2} \rho^{a}\,_{bc} H^{b}\,_{l} e^{l}
\wedge e^{c} + \frac{1}{2} \rho^{a}\,_{bc} H^{c}\,_{s} e^{b} \wedge e^{s}
\nonumber \\
- d \phi \wedge \tilde{\sigma}^{a} - \phi\,D \tilde{\sigma}^{a} + \phi
K_{cs} \rho_{n}\,^{ca} e^{s} \wedge e^{n} + \partial_{c} \phi \pi^{ca}
+ \lambda^{a}\,_{b} \rho^{b} = 0
\label{80}
\end{eqnarray}
where
\begin{equation}
{\cal L}_{\vec N} \rho^{a} = D ( N^{s} \iota_{s} \rho^{a} ) + N^{s}
\iota_{s} D \rho^{a} .
\label{81}
\end{equation}
Equation (\ref{80}) is the evolution equation for magnetic 2-form
$\rho^{a}$. In order to obtain the evolution equation for electric
1-form $\tilde{\sigma}^{a}$, we should act by $ \left(
\partial_{y} - {\cal L}_{\vec N} \right) $ to both sides of
equation (\ref{55}) and compare it with equation (\ref{75}). These
equations are complicated and will not be given here.
\section{Conclusion}
Using the language of differential forms we have presented the
gravitational field equations of 3-brane dynamics with the
Euler-Poincar\'{e} Lagrangian and a cosmological constant in
five-dimensions with $\mathbb{Z}_{2}$ symmetry. We have used a
more general ADM-type approach in which the normal to the brane
surface possesses an acceleration through the lapse function
$\phi$ and the shift vector $N^{a}$. In this approach effective
gravitational field equations on the brane obtained earlier in
\cite{maeda2} still remain to be the same, however in our case
they are given in terms of differential forms namely, either the
electric 1-form coming from the bulk Riemann curvature 2-form
$\mathbb{R}^{5a}$, or the conformal electric 1-form coming from
the bulk conformal curvature 2-form $\mathbb{C}^{5a}$. The field
equations involve the terms due to the cosmological constant as
well. We have also derived the evolution equations into the bulk
space which, unlike the on-brane field equations involve the terms
determined by the acceleration of the normals to the brane surface
in the ADM-type non-geodesic slicing of the bulk spacetime. In
this sense, the evolution off-brane equations  can be thought of
as more general ones and the consistent analysis of the non-linear
brane dynamics would certainly benefit from it. Therefore, the
developed formalism will be useful in studying the exact
gravitational solutions on the brane, like for instance black
holes and cosmological perturbations.

\section{Acknowledgement}

\noindent HC is supported through a Post-Doctoral Research
Fellowship by the Scientific and Technical Research Council of
Turkey (T\"{U}B\.{I}TAK). TD acknowledges partial support by the
Turkish Academy of Sciences (T\"{U}BA).

\end{document}